\documentstyle[aps,prb,twocolumn,epsf]{revtex}
\begin{document}
\draft
\wideabs{
\title {
Low thermal conductivity of the layered oxide (Na,Ca)Co$_2$O$_4$:
Another example of a phonon glass and an electron crystal
} 
\author{
K. Takahata, Y. Iguchi, D. Tanaka, and T. Itoh 
}
\address{
Department of Applied Physics, 
Waseda University, Tokyo 169-8555, Japan}
\author{ I. Terasaki\cite{byline}}
\address{
Department of Applied Physics, 
Waseda University, Tokyo 169-8555, Japan\\
and also Precursory Research for Embryonic Science and
Technology, Japan Science Technology, Tokyo 102-0074, Japan
}

\date{\today}
\maketitle
\begin{abstract}
The thermal conductivity of polycrystalline samples of 
(Na,Ca)Co$_2$O$_4$ 
is found to be unusually low, 20 mW/cmK at 280 K. 
On the assumption of the Wiedemann-Franz law,
the lattice thermal conductivity is
estimated to be 18 mW/cmK at 280 K,
and it does not change appreciably with the substitution
of Ca for Na. 
A quantitative analysis has revealed 
that the phonon mean free path is 
comparable with the lattice parameters,
where the point-defect scattering plays an important role.
Electronically the same samples show a metallic conduction 
down to 4.2 K,
which strongly suggests that NaCo$_2$O$_4$ exhibits 
a glass-like poor thermal conduction together with 
a metal-like good electrical conduction.
The present study further suggests that
a strongly correlated system with
layered structure can act as a material of
a phonon glass and an electron crystal.
\end{abstract}


} 


Thermoelectric  materials have recently
attracted a renewed interest as an application to
a clean energy-conversion system. \cite{Mahan} 
The conversion efficiency of a thermoelectric material is 
characterized by the figure of merit 
$Z = S^{2}/\rho\kappa$, 
where $S$, $\rho$ and $\kappa$ are the thermopower, 
the resistivity and the thermal conductivity, respectively. 
At a temperature $T$,
a dimensionless value of $ZT$ is required to be more than unity for
a good thermoelectric material,
which is, however, difficult to realize.
We have found a large thermopower
(100 $\mu$V/K at 300 K) and
a low resistivity (200 $\mu\Omega$cm at 300 K)
for NaCo$_2$O$_4$ single crystals. \cite{Terra} 
These parameters suggest that 
NaCo$_2$O$_4$ is a potential thermoelectric material. 
An important finding is that the
transport properties are difficult to understand 
in the framework of a conventional one-electron picture
based on band theories.
We have proposed that strong electron-electron 
correlation plays an important role 
in the enhancement of the thermopower. \cite{Terra,Itoh,Kawata}
Very recently Ando {\it et al}. \cite{Ando} have found that 
the electron specific-heat coefficient of NaCo$_2$O$_4$
is as large as 48 mJ/molK$^2$,
which is substantially enhanced from the free-electron value,
possibly owing to the strong correlation.

In addition to a large thermopower and a low resistivity,
a thermoelectric material is required to show a low thermal conductivity. 
In view of this, a filled skutterudite Ce(Fe,Co)$_4$Sb$_{12}$
shows quite interesting properties. \cite{Morelli,Sales1,Sales2,Keppens}
The most remarkable feature of this compound is that 
``filled'' Ce ions make the lattice thermal 
conductivity several times lower 
than that for an unfilled skutterudite CoSb$_3$. \cite{Morelli}
The Ce ions are weakly bound in an oversized atomic cage 
so that they will vibrate independently from the other atoms 
to cause large local vibrations. \cite{Keppens} 
This vibration and the atomic cage are named 
``rattling'' and a ``rattling site'', respectively.
As a result, the phonon mean free path
can be as short as the lattice  parameters.
Namely this compound has a poor thermal conduction like a glass 
and a good electric conduction like a crystal, 
which Slack \cite{Slack}
named a material of ``a phonon glass and an electron crystal''.
It should be mentioned that rattling is not the only reason
for the low thermal conductivity,
where point defects and/or solid solutions significantly
reduce the thermal conductivity
of La$_x$(Fe,Co)$_4$Sb$_{12}$ ($x<1$) \cite{Meisner}
and Co$_{1-x}M_x$Sb$_3$ ($M$=Fe, Ni, and Pt). \cite{Katsuyama,Anno}
Nevertheless a search for materials having rattling sites
is a recent trend for thermoelectric-material hunting.
Through this search,
BaGa$_{16}$Ge$_{30}$  \cite{Nolas} and
Tl$_2M$Te$_5$ ($M$=Sn and Ge) \cite{Sharp,Sales3} 
have been discovered as potential thermoelectric materials
with low thermal conductivity.

A preliminary study of the thermal conductivity of 
polycrystalline NaCo$_2$O$_4$,
which has {\it no rattling sites},
revealed a low value of
15-20 mW/cmK at 300 K. \cite{Yakabe}
This is indeed unexpectedly low,
because a material consisting of light atoms 
such as oxygens will have a high thermal conductivity.
In fact, polycrystalline samples of a high-temperature 
superconducting copper oxide show a higher value of 
40-50 mW/cmK at 300 K. \cite{Cohn,Ikebe}
This is qualitatively understood from its crystal structure
as schematically shown in the inset of Fig.~1.
NaCo$_2$O$_4$ is a layered oxide, which consists of
the alternate stack of the CoO$_2$ layer and the Na layer.
The CoO$_2$ layer is responsible for the electric conduction,
whereas the Na layer works only as a charge reservoir to
stabilize the crystal structure.
The most important feature is that the Na ions randomly occupy
50\% of the regular sites in the Na layer. 
The Na layer is highly disordered like an amorphous solid,
and it looks like a glass for the in-plane phonons.
Thus significant reduction of the thermal conductivity
is likely to occur in the sandwicth structure made of 
the crystalline metallic layers and 
the amorphous insulating layers. \cite{Terra2}

In this paper, we report on measurements and quantitative analyses on 
the thermal conductivity of polycrystalline samples of
(Na,Ca)Co$_2$O$_4$ from 15 to 280 K. 
The observed thermal conductivity is like that for a disordered crystal, 
and is insensitive to the substitution of Ca for Na.
These results imply that the phonon mean free path is
as short as the lattice parameters, 
and a semi-quantitative analysis reveals
that the point-defect scattering due to the solid solution
of Na ions and vacancies effectively reduces 
the lattice thermal conductivity down to 15-20 mW/cmK.  
On the other hand, the electrical resistivity remains 
metallic down to 4.2 K, which means that 
the electron mean free path is much longer than the lattice parameters.
Thus NaCo$_2$O$_4$ can be a material of 
a phonon glass and an electron crystal,
whose conduction mechanisms are qualitatively different 
from those of the ``rattler'' model of 
the filled skutterudite. \cite{Sales1,Sales2,Keppens}


Polycrystalline samples of 
Na$_{1.2-x}$Ca$_x$Co$_2$O$_4$ ($x$=0, 
0.05, 0.10 and 0.15) were prepared 
through a solid-state reaction. 
Starting powders of NaCO$_3$, CaCO$_3$ and Co$_3$O$_4$
were mixed and calcined at 860$^{\circ}$C for 12 h. 
The product was finely ground, pressed into a pellet, 
and sintered at 920$^{\circ}$C for 12 h. 
Since Na tends to evaporate during calcination, 
we added 20 \% excess Na. 
Namely we expected samples of the nominal 
composition of Na$_{1.2-x}$Ca$_x$Co$_2$O$_4$ to be 
Na$_{1-x}$Ca$_x$Co$_2$O$_4$,
which we will denote as (Na,Ca)Co$_2$O$_4$.

The thermal conductivity was measured 
using a steady-state technique
in a closed refrigerator pumped down to 10$^{-6}$ Torr.
The sample was pasted on a copper block with silver paint
(Dupont 4922) to make a good thermal contact 
with a heat bath, and on the other side of the 
sample a chip resistance heater (120 $\Omega$) 
was pasted to supply heat current. 
Temperature gradient was monitored 
by a differential thermocouple made of Chromel-Constantan, 
while temperature was monitored with  
a resistance thermometer (Lakeshore CERNOX 1050). 


Figure 1 shows the thermal conductivity of (Na,Ca)Co$_2$O$_4$.
The substitution of Ca for Na only slightly decreases
the thermal conductivities of (Na,Ca)Co$_2$O$_4$.
This makes a remarkable contrast to the change of the resistivity 
with the Ca substitution. \cite{Itoh,Kawata} 
The magnitude (20 mW/cmK at 280 K) 
is as low as that of a conventional thermoelectric material
such as Bi$_2$Te$_3$, \cite{Caillat}
which is consistent with the previous study. \cite{Yakabe}

Let us make a rough estimate of 
the phonon mean free path ($\ell_{ph}$)
for NaCo$_2$O$_4$ at 280 K.
In the lowest order approximation, 
the lattice thermal conductivity $\kappa_{ph}$ 
is expressed by \cite{AM}
$$ \kappa_{ph}=\frac{1}{3}cv\ell_{ph},$$
where $c$ and $v$ are the lattice specific heat and 
the sound velocity.
Since we consider a moderately high temperature region 
where phonons are sufficiently excited, 
we assume $c=3Nk_B$ 
($N$ is the number of atoms per unit volume).
The sound velocity is associated with 
the Debye temperature $\theta_D$ as
$$ \theta_D
  =\frac{\hbar v}{k_B}(6\pi^2N)^{\frac{1}{3}}.$$
We employ $\theta_D$=350 K from the recent 
specific-heat data, \cite{Ando}
and get $\ell_{ph}$ = 6.7 \AA~for 20 mW/cmK,
which is comparable with the in-plane lattice parameter (3 \AA).
This picture is intuitively understood from the fact that
the Na layer is highly disordered. 
Note that the observed data of 20 mW/cmK includes the 
electron thermal conductivity, and thus 
the obtained value of 6.7 \AA~ gives the upper limit of
the phonon mean free path.  

Figure 2 summarizes the thermoelectric parameters of 
NaCo$_2$O$_4$.
In Fig. 2(a) are shown the thermal conductivity
(the same data as $x$=0 in Fig. 1) and the figure of merit
calculated using the resistivity and the thermopower
of the same sample.
We also plot the electron thermal conductivity ($\kappa_{el}$) 
estimated from the resistivity
on the assumption of the Wiedemann-Franz law
as $\kappa_{el}=L_0T/\rho$ 
($L_0=\pi^2k_B^2/3e^2$ is the Lorentz number).
$\kappa_{el}$ is 10\% of $\kappa$, and 
the heat conduction is  mainly determined by the phonons.
The figure of merit is 10$^{-4}$K$^{-1}$ above 100 K, 
which is largest among oxides, \cite{Mahan}
but does not yet reach the criteria of $ZT=1$.
Much progress is thus needed to realize oxide thermoeletcrics.

In Fig. 2(b), the resistivity and the thermopower are plotted 
as a function of temperature,
which reproduce the pioneering work on the Na-Co-O system 
by Molenda {\it et al.} \cite{Molenda}
The temperature dependence of the resistivity
is essentially the same as that for
the in-plane resistivity of the single crystals,
though the magnitude is much higher owing to the grain-bondary scattering.
It should be noted that the resistivity exhibits metallic conduction 
down to 4.2 K without any indication of the localization.
This implies that the electron mean free path is 
much longer than the lattice parameters.
Previously we showed that the electron mean free path 
of the single crystal is as long as 230 \AA~ at 4.2 K 
along the in-plane direction. \cite{Terra}
We can therefore say
that the phonon mean free path 
is much shorter than the electron mean free path.
This is nothing but a material of
a phonon glass and an electron crystal. \cite{Slack}

Here we will compare the measured thermal conductivity with 
the phonon-scattering theory by Callaway. \cite{Callaway1,Callaway2}
The total scattering rate $\tau^{-1}$ is given 
as the sum of three scattering rates as   
\begin{eqnarray*}
 \tau^{-1} &= &{\tau_{pd}}^{-1}+{\tau_{ph-ph}}^{-1}+{\tau_0}^{-1}\\
           &= &A\omega^4  +B\omega^2 + v/L
\end{eqnarray*}
where ${\tau_{pd}}^{-1}$, ${\tau_{ph-ph}}^{-1}$ and ${\tau_0}^{-1}$
are the scattering rates for the point-defect scattering,
the phonon-phonon scattering, and the boundary scattering, respectively.
For a phonon frequency $\omega$, the three scattering rates 
are written as $A\omega^4$, $B\omega^2$ and $v/L$,
where $A$, $B$ and $L$ are characteristic parameters.
According to Ref. \onlinecite{Klemens}, $A$ is expressed as 
$A=\Omega_0\sum f_i(1-M_i/M)^2/4\pi {v}^3$,
where $\Omega_0$ is the unit cell volume,
$M_i$ is the mass of an atom,
$f_i$ is the fraction of an atom with mass $M_i$,
and $M=\sum f_iM_i$ is the average mass.
We calculated $A$ for (Na,Ca)Co$_2$O$_4$
by following the method in Ref. \onlinecite{Meisner},
where Na (23g/mol), Ca (40g/mol) and $\Box$ (vacancy)
make a solid solution in the ratio of 
Na:Ca:$\Box$=(1-$x$):$x$:1.
$B$ is a temperature-dependent parameter,
which is proportional to $T$ at high temperatures ($B\sim CT$).
It should be noted that the phonon-phonon scattering gives
$\kappa \propto 1/\sqrt{ACT}$ at high temperatures
in the presence of a large $A$. \cite{Callaway2}
As clearly shown in Fig. 1,
$\kappa$ for (Na,Ca)Co$_2$O$_4$ increases with $T$, 
implying that the phonon-phonon scattering is neglegibly small. 
Thus $L$ corresponding to an inelastic scattering length
is the only fitting parameter.

In Fig. 3, the measured $\kappa_{ph}$ $(= \kappa-\kappa_{el})$ 
of NaCo$_2$O$_4$ is compared with two theoretical curves.
Sample \#1 is the same sample as shown in Fig. 1, 
and Sample \#2 is another sample prepared in a different run.
Curve A is the calculation using the phonon-scattering 
theory, \cite{Callaway1,Callaway2}
where $L$=0.2 $\mu$m is used.
As expected, the point-defect scattering 
quite effectively reduces the thermal conductivity 
by two or three orders of magnitude.
The Ca substitution effect is also consistently 
explained as shown in the inset,
where data points (as indicated by open circles)
in different runs are added to show the reproducibility.
Although the solid solution of Na and $\Box$ dominates
$\kappa_{ph}$, the theory predicts  a small correction due to the 
substitution of Ca (as indicated by the solid line), 
which is in good agreement with the observation.
This directly indicates that the point-defect scattering
plays an important role in reducing $\kappa_{ph}$. 
A problem is the physical meaning of $L$=0.2 $\mu$m:
It is much longer than the electron mean free path (10-10$^2$ \AA), 
but much shorter than the grain size (10 $\mu$m).
Possible candidates are the average distance
of stacking faults and/or interlayer disorder. 

The absence of the phonon-phonon scattering 
means that the phonon lifetime is extremely short, \cite{Comment}
and is rather characteristic of the thermal conductivity of a glass.
Curve B is the calculation of the minimum thermal conductivity 
$\kappa_{min}$ by Cahill {\it et al.}, \cite{Cahill}
which has been compared with $\kappa_{ph}$ for a glass.
Although the calculated $\kappa_{min}$ is one order of magnitude smaller
than the measured $\kappa_{ph}$,
such a deviation is also seen in other disordered crystals.
Note that $\kappa_{ph}\propto T^3$ is not seen 
for NaCo$_2$O$_4$ at low temperatures,  
which is a hallmark of disordered crystals.
Since $\kappa_{ph}\propto T^3$ is usually seen below 10 K,
this is possibly because the measurement temperature was high.
NaCo$_2$O$_4$ consists of the sandwich structure of
the amorphous and crystalline layers, and the heat conduction process
is perhaps in between that for a mixed crystal and an amorphous solid.
Thus it should be further explored which curve is more likely
to capture the essential feature of the heat conduction
in NaCo$_2$O$_4$.

We propose that a layered material consisting of
a strongly correlated conducting layer and
disordered insulating layer
can be a promising thermoelectric material. 
If the heavy-fermion system is realized in the strongly correlated layer,
$S^2/\rho$ can be increased through the mass enhancement 
due to the spin fluctuation. \cite{Mahan,Webb}
Recently we proposed that the effective mass of NaCo$_2$O$_4$
is enhanced as much as that for CePd$_3$. \cite{Terra2}
Meanwhile, in the disordered insulating layer,
the lattice thermal conductivity can be
minimized by the disorder that causes little effect on 
the electric conduction.
In this context, it will work as 
a material of a phonon glass and an electron crystal.

This scenario might be compared with  
the thermoelectric superlattices extensively studied 
by Dresselhaus et al. \cite{Hicks,Dresselhaus}
At present, no band calculation of NaCo$_2$O$_4$ is available, 
but the band calculation of isostructural LiCoO$_2$ shows
that the valence bands do not show any sub-band structure
expected from the 2D quantum confinement. \cite{Aydinol}
This means that the electronic states of NaCo$_2$O$_4$ 
are not very anisotropic in a one-electron picture.
This situation is essentially identical to 
the band picture of high-temperature superconductors.
We think that the enhancement of the thermopower of NaCo$_2$O$_4$
should not be attributed to the quantum confinement of
the semiconductor superlattices,
but to the strong correlation.


In summary, we prepared polycrystalline samples of (Na,Ca)Co$_2$O$_4$ 
and measured the thermal conductivity from 15 to 280 K.
We have found that the phonon mean free path is 6.7 \AA~ at 280 K,
which is much shorter than the electron mean free path.
This means that (Na,Ca)Co$_2$O$_4$ acts as a materials of a phonon glass
and an electron crystal, though it has no rattling sites.
We have compared the experimental data with the phonon-scattering 
theory and the minimum thermal conductivity,
and have found that the point-defect scattering plays an important role.


The authors would like to appreciate J. Takeya and Y. Ando
for technical help for thermal-conductivity measurement.
They also thank T. Kawata, T. Kitajima, T. Takayanagi,
T. Takemura, and T. Nonaka for collaboration.



\begin{figure}
\centerline{\epsfxsize=7cm \epsfbox{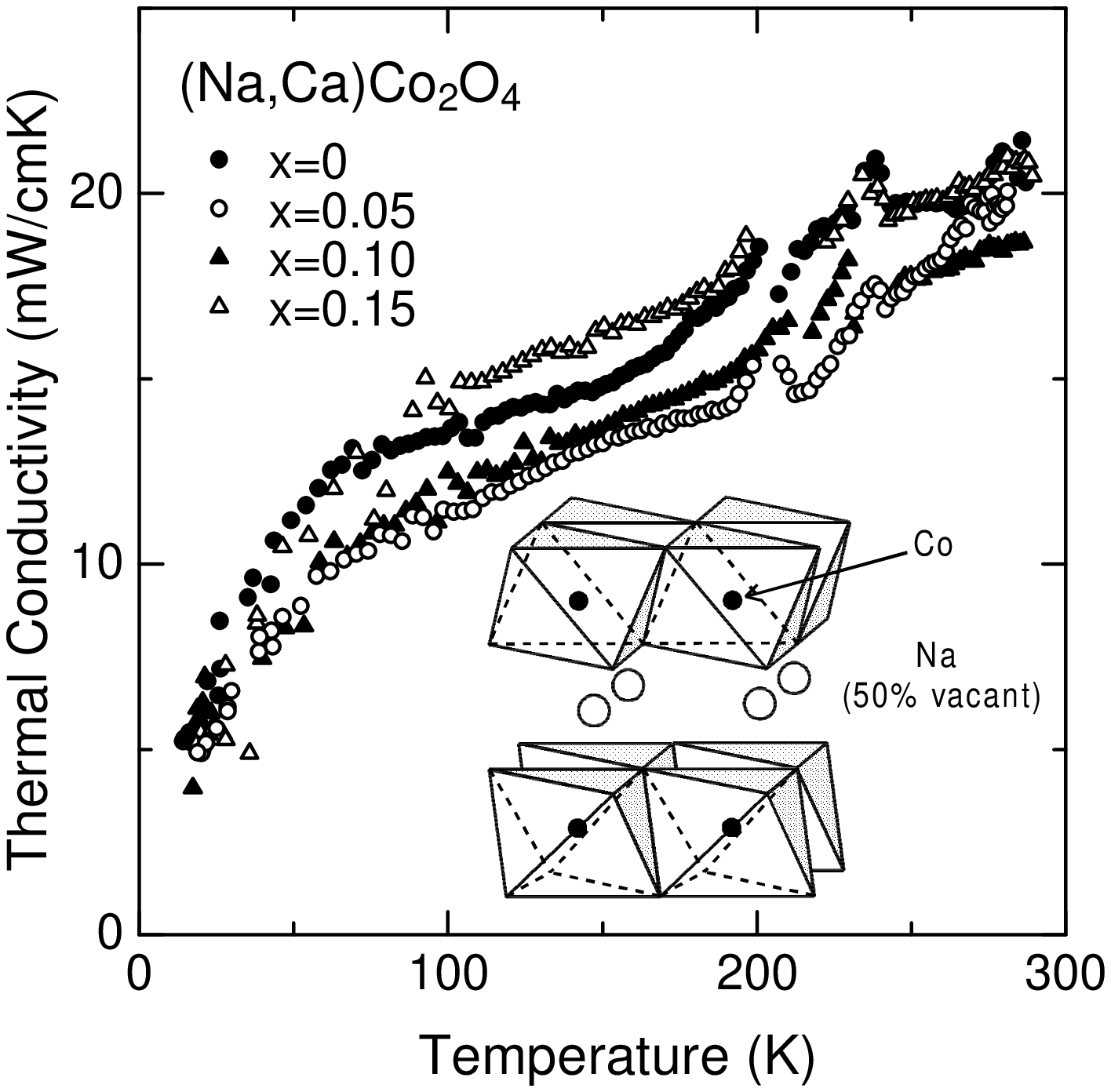}}
\caption{%
The thermal conductivity of (Na,Ca)Co$_2$O$_4$ plotted as 
a function of temperature.
The inset shows the schematic picture of 
the crystal structure of NaCo$_2$O$_4$.
}
\end{figure}

\begin{figure}
\centerline{\epsfxsize=7cm \epsfbox{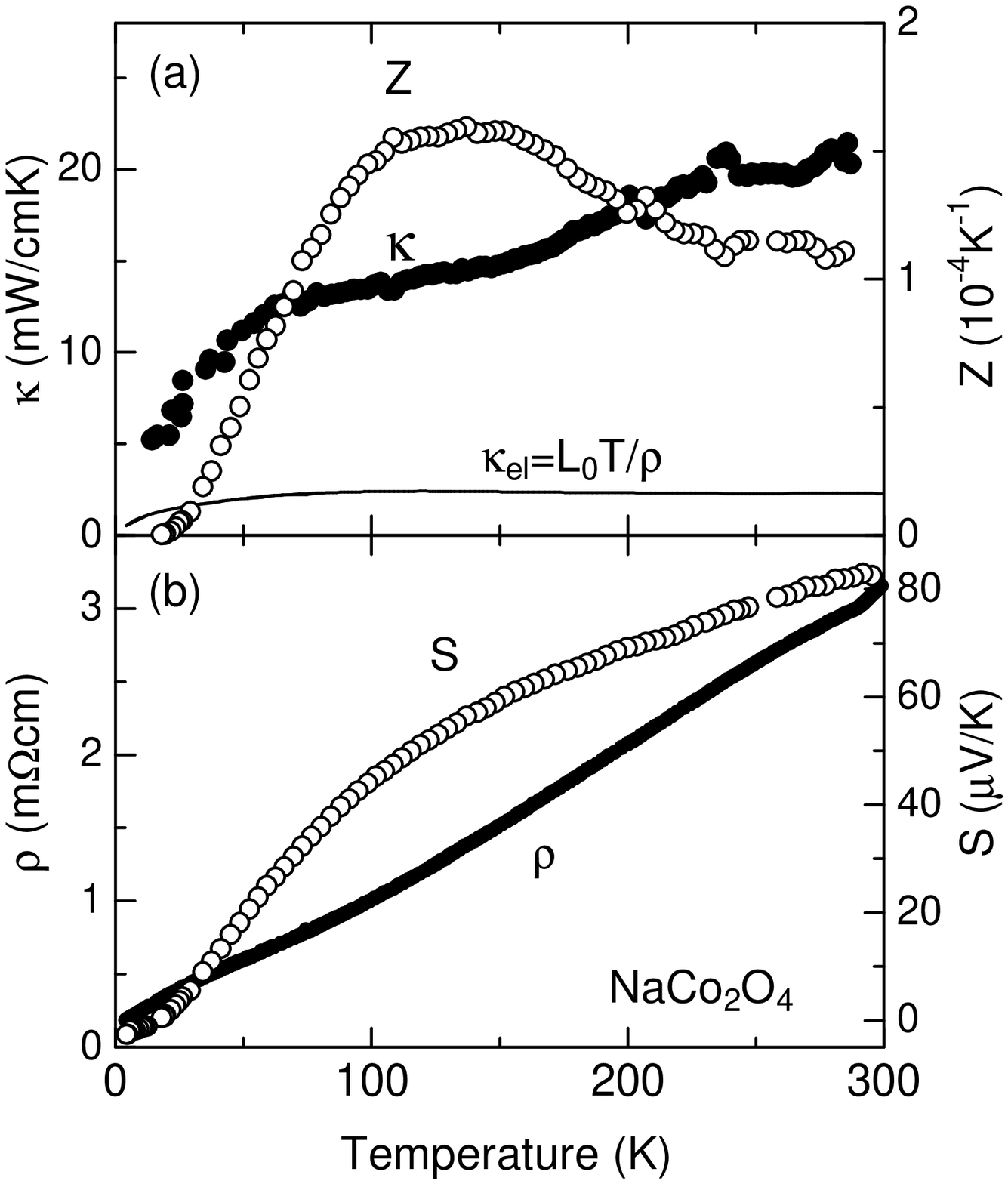}}
\caption{
 The thermoelectric parameters of polycrystalline NaCo$_2$O$_4$.
 (a) The thermal conductivity ($\kappa$) and the figure of merit ($Z$);
 (b) The resistivity ($\rho$) and the thermopower ($S$).
 Note that the electron thermal conductivity ($\kappa_{el}$) evaluated
 through the Wiedemann-Franz law is shown by the solid curve,
 where $L_0$ is the Lorentz number ($=\pi^2k_B^2/3e^2$).
 }
\end{figure}

\begin{figure}
\centerline{\epsfxsize=7cm \epsfbox{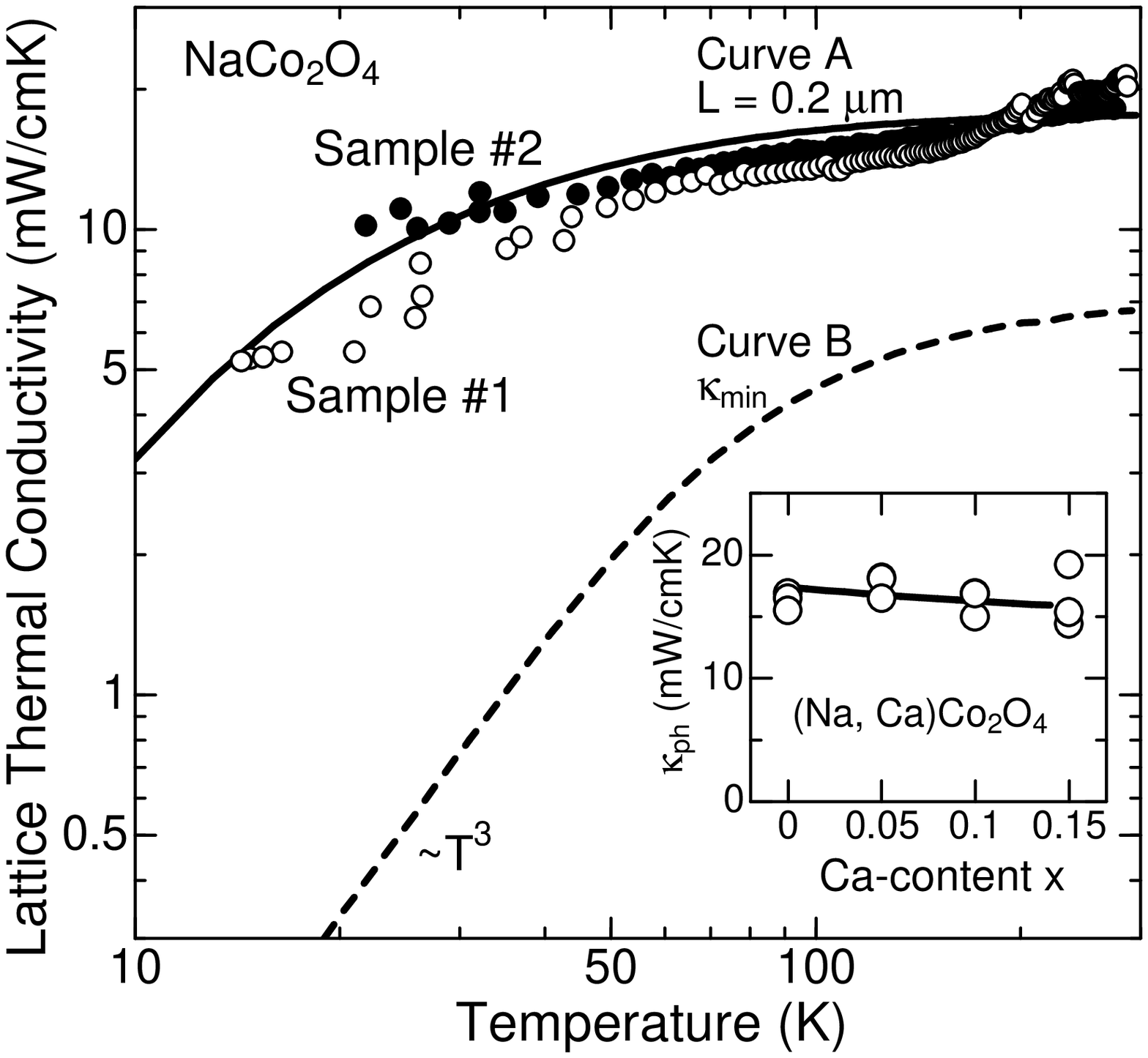}}
\caption{%
The lattice thermal conductivity ($\kappa_{ph}$) of NaCo$_2$O$_4$.
The open and closed circles represent $\kappa_{ph}$ measured 
for Sample \#1 and \#2, where the electron thermal conductivity
was estimated through the Wiedemann-Franz law.
Curve A is the calculation proposed 
by Callaway. \protect{\cite{Callaway1,Callaway2}}
Curve B is the minimum thermal conductivity
proposed by Cahill {\it et al.} \protect{\cite{Cahill}} 
The inset shows $\kappa_{ph}$ of (Na,Ca)Co$_2$O$_4$ at 200 K.
The solid line is the same calculation as Curve A.   
}
\end{figure}


\begin{thebibliography}{99}
\bibitem{Comment}
 Another possibility is that the phonon in NaCo$_2$O$_4$
 is quite harmonic.
 In this case, there are little disspation channels,
 and $\kappa_{ph}$ would be much higher.
\end{thebibliography}
\end{document}